\documentclass[a4paper]{jpconf}
\usepackage{graphicx}
\newcommand{\be}{\begin{equation}}
\newcommand{\ee}{\end{equation}}
\newcommand{\ba}{\begin{eqnarray}}
\newcommand{\ea}{\end{eqnarray}}
\begin{document}
\title{An A4 model for neutrinos}

\author{Eduardo Peinado}

\address{AHEP Group, Institut de F\'{\i}sica Corpuscular --
C.S.I.C./Universitat de Val{\`e}ncia \\
\small Edificio Institutos de Paterna, Apt 22085, E--46071 Valencia, Spain}

\ead{epeinado@ific.uv.es}

\begin{abstract}
The study of an extension of the standard model based on the flavor 
symmetry $A_4$ is presented.
Neutrino Majorana mass terms arise from dimension five operator and
charged lepton masses from renormalizable Yukawa couplings. 
We introduce three Higgs doublets that belong to one triplet irreducible
representation of $A_4$. We study the most general $A_4$-invariant 
scalar potential and the phenomenological consequences of the model. 
We find that the reactor angle could be as large as, $\sin^2\theta_{13_{\mbox{max}}}\sim 0.03$, while the atmospheric mixing angle $\theta_{23}$ is close to maximal, $\sin^2 \theta_{23}=1/2$.\end{abstract}

\section{\label{int}Introduction}
Experiments on neutrino oscillations confirmed that neutrinos are massive and mix among themselves like the quarks do. In contrast to the quark sector, neutrino oscillations have two large mixing angles~\cite{Schwetz:2008er}.  
In the literature continuous as well as 
discrete (Abelian and non-Abelian) flavor symmetries have been extensively studied. However so far we do not have a unique top-down hint for the choice of the
flavor symmetry. In a bottom-up approach simplicity and predictivity are possible criteria that we can use. In the quark sector the heaviness of the top quark suggests that first and second families could belong to a doublet irreducible representation and the third family belongs to a singlet representation of the flavor symmetry group.
Differently the large solar and atmospheric neutrino mixing angles suggest
that the three neutrino families belong to one triplet irreducible representation of a flavor group. The group of even permutation of four objects, $A_4$\cite{Babu:2002dz,Altarelli:2005yp}, is the smallest non-Abelian discrete group with triplet irreducible representation. 
%Here we study a model based on $A_4$. 
Most of the models based on $A_4$ need to introduce extra auxiliar Abelian 
symmetries and supersymmetry  (or extra dimensions) in order to reproduce
tri-bimaximal mixing~\cite{Harrison:2002er}. %There are also extension of the SM based on continuous flavor symmetries~\cite{continuous}
 
In this paper we  use the $A_4$ symmetry group as the flavor group but we renounce to predict the tri-bimaximal mixing. 
The Dirac mass matrices arise from renormalizable operator coupled with
the Higgs, that is the Yukawa interactions. Neutrino Majorana mass terms
are generated from a dimension five Weinberg operator.
It is a known fact that $A_4$ can be broken spontaneously into its $Z_3$ or $Z_2$ subgroups assuming the vacuum expectation values (vevs) to be real \footnote{When $A_4$ is broken into $Z_3$ in the charged lepton sector and into $Z_2$ in the neutrino sector the lepton mixing is  tri-bimaximal.}. Recently  in \cite{Lavoura:2007dw}, a model for quarks mixing has been suggested where the most generic
$A_4$-invariant potential has complex solutions. In this case $A_4$ is completely broken. 
This solution open new possibilities for the description of the leptonic sector that deserve further investigation. 
%while the implications in the quark sector  has been studied in~\cite{Lavoura:2007dw}.

In the next section we present the model, in section \ref{cl} and \ref{neu} we discuss the charged lepton and neutrino 
mass matrices respectively. In section \ref{phen} we discuss the implication of the model and finally, in section \ref{conc} we summarize the results of the model.

\section{The Model}
\label{themodel}

In our model~\cite{Morisi:2009sc} the Higgs sector is extended from one $SU(2)_L$-doublet to three $SU(2)_L$-doublets belonging to a triplet irreducible representation of $A_4$. The left-handed as well as the right-handed charged leptons, belong to the triplet irreducible representation of $A_4$. 
The irreducible representation assignment for the particles is given in Table \ref{tab:Multiplet1}.
\begin{table}[h!]
\begin{center}
  \begin{tabular}{|l||ll||l|}
\hline
fields & $L_i$ & $l^c_i$ & $\phi_{i}$  \\
\hline
$SU(2)_L$ & 2& 1& 2 \\
$A_4$  &  3& 3   & 3 \\
\hline
\end{tabular}
\caption{Lepton multiplet structure of the model}
\label{tab:Multiplet1}
\end{center}
\end{table}

%$A4$ can be generated by two basic permutations $S$ and $T$. If we work in the basis where $S$ is diagonal~\cite{Altarelli:2007cd}, 
$A_4$ is the group of even permutations of $4$ objects and is isomorphic to the symmetries of the tetrahedron. $A_4$ can be generated by two generators $S$ and $T$ with the properties
\be
S^2=T^2=(ST)^3=1.
\ee
$A_4$ contain one 3-dimensional representation, ${3}$, and three one-dimensional, ${1}$, ${1}^{\prime}$  and ${1}^{\prime\prime}$. The product of two 3 gives ${ 3} \otimes { 3} = { 1} \oplus { 1}^{\prime} \oplus  { 1}^{\prime \prime} \oplus { 3} \oplus { 3}$ and ${ 1}^{\prime} \otimes { 1}^{\prime} = { 1}^{\prime \prime}$, ${ 1}^{\prime} \otimes { 1}^{\prime \prime} = { 1}$, ${ 1}^{\prime \prime} \otimes { 1}^{\prime \prime} = { 1}^{\prime}$ etc. 
For two triplets $3_a\sim (a_1,a_2,a_3)$, $3_b\sim (b_1,b_2,b_3)$ the irreducible representations obtained from their product are:
\begin{equation}
1=a_1b_1+a_2b_2+a_3b_3,
\end{equation}
\begin{equation}
1'=a_1b_1+\omega^2 a_2b_2+\omega a_3b_3,
\end{equation}
\begin{equation}
1"=a_1b_1+\omega a_2b_2+\omega^2 a_3b_3,
\end{equation}
\begin{equation}
3\sim (a_2b_3, a_3b_1, a_1b_2),
\end{equation}
\begin{equation}
3\sim (a_3b_2, a_1b_3, a_2b_1),
\end{equation}
this in the basis of $S$ diagonal and where $\omega=e^{i2\pi/3}$.

The most general renormalizable Yukawa Lagrangian for the charged leptons in the model is
\be
\begin{array}{lll}
L_{\mbox{Yukawa}}&=&y_1\left( \bar{L_1}\phi_{3}\l^c_2 + \bar{L_2}\phi_{1}\l^c_3 +  \bar{L_3}\phi_{2}\l^c_1  \right)+ \\ 
&&+y_2\left( \bar{L_1}\phi_{2}\l^c_3 + \bar{L_2}\phi_{3}\l^c_1 +  \bar{L_3}\phi_{1}\l^c_2  \right) .
\end{array}
\ee
Once the electroweak symmetry (EW) is broken, the charged lepton mass matrix are obtained from this Yukawa Lagrangian:
\be
M_{l}=\left(
\begin{array}{ccc}
0 & y_1 \langle\phi_3 \rangle & y_2 \langle\phi_2 \rangle \\
y_2 \langle\phi_3 \rangle & 0 & y_1 \langle\phi_1 \rangle \\
y_1 \langle\phi_2 \rangle & y_2 \langle\phi_1 \rangle & 0
\end{array}
\right).
\label{me}
\ee
The most general neutrino dimension five operator invariant under $A_4$ (see for instance~\cite{Morisi:2009sz}), is given by
\be
\begin{array}{lll}
\mathcal{L}_{5d}&=&\beta \left(LL\right)_3\left(HH\right)_3+k\left(LL\right)_1\left(HH\right)_1+\alpha^{\prime}\left(LL\right)_{1^{\prime}}\left(HH\right)_{1^{\prime\prime}}+\alpha^{\prime\prime}\left(LL\right)_{1^{\prime\prime}}\left(HH\right)_{1^{\prime}}+ \\ \\ &+&
\left[
a\left(LH\right)_{3^a}\left(LH\right)_{3^a}+b\left(LH\right)_{3^a}\left(LH\right)_{3^b}+c\left(LH\right)_{3^b}\left(LH\right)_{3^a}+d\left(LH\right)_{3^b}\left(LH\right)_{3^b}\right]+  \\ \\
&+& l\left(LH\right)_1\left(LH\right)_1+l^{\prime}\left[\left(LH\right)_{1^{\prime}}\left(LH\right)_{1^{\prime\prime}}+\left(LH\right)_{1^{\prime\prime}}\left(LH\right)_{1^{\prime}}\right],
\label{yn}
\end{array}
\ee
where $\beta,k,\alpha,\alpha', \alpha'', a,b,c,d$ are arbitrary complex couplings.
Once the EW symmetry is broken, the Majorana neutrino mass matrix is given by
\be
M_\nu=\left(\begin{array}{ccc}
x \langle \phi_1\rangle^2+y \langle \phi_2\rangle^2+z \langle \phi_3\rangle^2 & \kappa \langle \phi_1\rangle \langle \phi_2\rangle & \kappa \langle \phi_1\rangle \langle \phi_3\rangle \\
\kappa \langle \phi_1\rangle \langle \phi_2\rangle & z \langle \phi_1\rangle^2+x \langle \phi_2\rangle^2+y \langle \phi_3\rangle^2 & \kappa \langle \phi_2\rangle \langle \phi_3\rangle \\
\kappa \langle \phi_1\rangle \langle \phi_3\rangle & \kappa \langle \phi_2\rangle \langle \phi_3\rangle & y \langle \phi_1\rangle^2+z \langle \phi_2\rangle^2+x \langle \phi_3\rangle^2
\end{array}
\right),
\label{mnu}
\ee
where $x$, $y$, $z$ and $\kappa$ are functions of the couplings in eq. (\ref{yn}).

%%%%%%%%%%%%%

The most general renormalizable scalar potential
invariant under the symmetry $A_4$~\cite{Lavoura:2007dw,Morisi:2009sc} has the solution
 
\be
\left\langle  \phi_1 \right\rangle
= v_1 ,
\quad
\left\langle \phi_2 \right\rangle
= v e^{i \alpha / 2},
\quad
\left\langle  \phi_3 \right\rangle
= ve^{- i \alpha / 2}.
\label{vevs}
\ee

\section{Charged leptons}
\label{cl}
%If we assume the charged lepton assigned as follows:
%
%\[
%\begin{array}{ccc}
%L\sim 3& \mbox{and} & l_R \sim 3
%\end{array}
%\]

%We show in the appendix \ref{potential} that the minimization of the most general
%scalar potential for the Higgs doublet belonging to a triplet irreducible representation of $A_4$ gives\footnote{We choose a different solution from that used in ~\cite{Lavoura:2007dw} because we are interested in the breakdown of the $\mu-\tau$ symmetry.}

With the vevs in eq. (\ref{vevs}), the charged lepton mass matrix in eq. (\ref{me}) takes the form
\begin{equation}
M_{l}=
\left(
\begin{array}{ccc}
0 & a e^{i\alpha} & b e^{-i\alpha} \\
b e^{i \alpha} & 0 & a r \\
a e^{-i \alpha} & b r & 0
\end{array}
\right),
\label{me2}
\end{equation}
with the parameters in the matrix (\ref{me2}) defined by $a=y_{1}v$, $b=y_{2}v$ and $r=v_{1}/v$. We write the symmetric matrix, $M_{l}M_{l}^{T}$ 
\begin{equation}
M_{l}M_{l}^{T}=\left(\begin{array}{ccc}
a^2+b^2 & abr &abr \\
abr & b^2+a^2r^2 & ab \\
abr & ab & a^2+b^2r^2
\end{array}
\right),
\end{equation}
which is diagonalized by an orthogonal matrix $O_{l}$. 
It is straightforward to obtain the analytical expressions for $a$, $b$ and $r$ as function of the charged lepton masses~\cite{Morisi:2009sc}, it can be written as
\be
\begin{array}{l}
r\approx \frac{m_{\tau}}{\sqrt{m_{e}m_{\mu}}}\sqrt{1-\frac{m_{e}^2m_{\mu}^2}{m_{\tau}^4}}, \\
a\approx \frac{m_{\mu}}{m_{\tau}}\sqrt{m_{e}m_{\mu}}\left[1+\frac{1}{2}\frac{m_{\mu}^2}{m_{\tau}^2}\right],\\
b\approx \sqrt{m_{e}m_{\mu}}\left[1-\frac{1}{2}\frac{m_{\mu}^2}{m_{\tau}^2}\right].
\end{array}\label{sollep}
\ee
With
\[
\begin{array}{ccc}
m_{e}=0.511006\, ~\mbox{MeV},& m_{\mu}=105.656\, ~\mbox{MeV},& m_{\tau}=1776.96\, ~\mbox{MeV},
\end{array}
\]
we have 
$a=0.43474$, $b=7.3471$ and $r=241.8582$. Note that $a<b\ll r$
thus the orthogonal matrix $O_l$ diagonalizing $M_l M_l^T$ is approximatively
\begin{equation}
O_{l_{12}}\approx \frac{b}{a} r^{-1}, \quad
O_{l_{13}}\approx \frac{a}{b} r^{-1}, \quad
O_{l_{23}}\approx \frac{a}{b} r^{-2}.
\end{equation}
The element, $O_{l_{12}}$,  give a contribution to the reactor mixing angle, $\theta_{13}$, see section \ref{nma}. The analytical expression for this element is given as
\be\label{ol12}
O_{l_{12}}\approx  \sqrt{\frac{m_{e}}{m_{\mu}}}\left[1-\left(\frac{m_{\mu}}{m_{\tau}}\right)^2\right].
\ee
The numerical expression for the matrix $O_l$ is 
\begin{equation}\label{Ol}
\begin{array}{lll}
O_{l}&=&\left(
\begin{array}{ccc}
 0.997 &0.069& 2.44\times 10^{-4} \\
-0.069  &  0.997  & 1.075 \times 10^{-6} \\
-2.439 \times 10^{-4} & -1.800 \times 10^{-5}  & 0.999
\end{array}
\right).
\end{array}
\end{equation}

\section{Neutrinos}
\label{neu}
The mass matrix for the neutrinos in eq. (\ref{mnu}) with the vevs in eq. (\ref{vevs}) takes the form
\be
M_\nu=\left(\begin{array}{ccc}
x r^2+y e^{-2 i \alpha}+z e^{2 i \alpha} & \kappa r e^{-i\alpha} & \kappa r e^{i\alpha}\\
\kappa r e^{-i\alpha} & z r^2+x e^{-2 i \alpha}+y e^{2 i \alpha} & \kappa  \\
\kappa r e^{i\alpha}& \kappa  & y r^2+z e^{-2 i \alpha}+x e^{2 i \alpha} 
\end{array}
\right).
\label{mnu20}
\ee
From the charged lepton sector we know that $r$ is fixed as shown in eq.\,(\ref{sollep}), and  $r\gg 1$, then we can neglect in the diagonal the terms not proportional to $r^2$. With this the mass matrix in eq. (\ref{mnu20}) can be written as
\be
M_\nu=\left(\begin{array}{ccc}
x r^2 & \kappa r e^{-i\alpha} & \kappa r e^{i\alpha}\\
\kappa r e^{-i\alpha} & z r^2 & \kappa  \\
\kappa r e^{i\alpha}& \kappa  & y r^2 
\end{array}
\right).
\label{mnu2}
\ee 
Note that there are 4 complex free parameters, $x,\,y,\,z,\,\kappa$ and one extra phase $\alpha$ coming from the Higgs sector. We can absorb
two phases in the fields, then it remains 7 free parameters in the neutrino mass matrix. Neutrino oscillation experiments determine two mass square difference
 $\Delta m_{12}^2\equiv m_2^2-m_1^2$ and $\Delta m_{13}^2\equiv |m_3^2-m_1^2|$ with the corresponding three mixing angle~\cite{Schwetz:2008er}.
If $\theta_{13}$ is different from zero, the Dirac phases could be probed in future experiments \cite{Ardellier:2006mn}.
The absolute neutrino mass scale can be probed in future tritium beta decay 
\cite{Drexlin:2005zt}
and neutrinoless double beta decay~\cite{mbbd} experiments.
While it will be hard to measure the two Majorana phase. There are seven measurable physical observable
plus two Majorana physical phases.

%Therefore in the general case the number of free parameters is equal to the number of experimental
%observables. However we can make some {\it ad hoc} assumptions reducing the number of free parameters as below.

\section{Phenomenology of the model}
\label{phen}
In our Model we have six free parameters, $x,~y,~\theta,~\phi_{xy}, \,\delta$ and $\alpha$ in the neutrino sector for nine physical parameters, $\Delta m_{12}^2$, $\Delta m_{13}^2$, $m_{ee}$, three mixing angles and two Majorana phases and the Jarlskog invariant $J$, for Dirac CP violation in the neutrino sector. We can construct the expressions for $\Delta m_{12}^2$ and $\Delta m_{13}^2$ and find $x$ and $y$ as functions of the observables, $\Delta m_{12}^2$, $\Delta m_{13}^2$, the mixing parameter $\theta$ and its relative phase $\phi_{xy}$.
This model is only compatible with inverted hierarchy neutrino mass spectrum ~\cite{Morisi:2009sc}. 

\noindent In the next subsections we present the predictions for the allowed region for mixing angles, the Jarlskog invariant as well as the neutrinoless double beta decay.

\subsection{Neutrino mixing angles}
\label{nma}
Recently has been given an indication that the reactor neutrino angle $\theta_{13}$
could be different from zero \cite{Schwetz:2008er}
\be\label{fog}
\sin^2\theta_{13}=0.016 \pm 0.010\, (1\sigma).
\ee

\noindent The $\theta_{13}^\nu$  angle coming from the diagonalization of the neutrino mass matrix,
is exactly zero in the limit $\delta, \alpha=0$.
However the reactor angle resulting from the product of the unitary
matrix that diagonalize the charged lepton matrix, eqs.\,(\ref{ol12}) and (\ref{Ol})
and the neutrino mass matrix, is different from zero and we have
\be\label{al0}
\sin^2\theta_{13}\approx\frac{m_e}{2 m_\mu}\approx 0.0024.
\ee

The solar mixing angle is given by the $\theta$ parameter up to corrections coming from the charged lepton sector of the order ${\cal O}(\sqrt{m_e/m_\mu})$. The parameters $\alpha$ and $\delta$ are related with the deviations of $\theta_{13}$ and $\theta_{23}$ from the zero and maximal values respectively.

%\noindent 

%For $\delta, \alpha\ne 0$ we have small deviation of 
%$\theta_{13}^\nu$ from zero. %We found that the corrections for $\theta_{13}$ coming from $\delta$ are smaller than the corrections coming from $\alpha$.  
%Here we study the case with $\alpha, \delta \ne0$.

\begin{figure}[h!]
%
%\hspace{0.5cm}
\includegraphics[width=8cm]{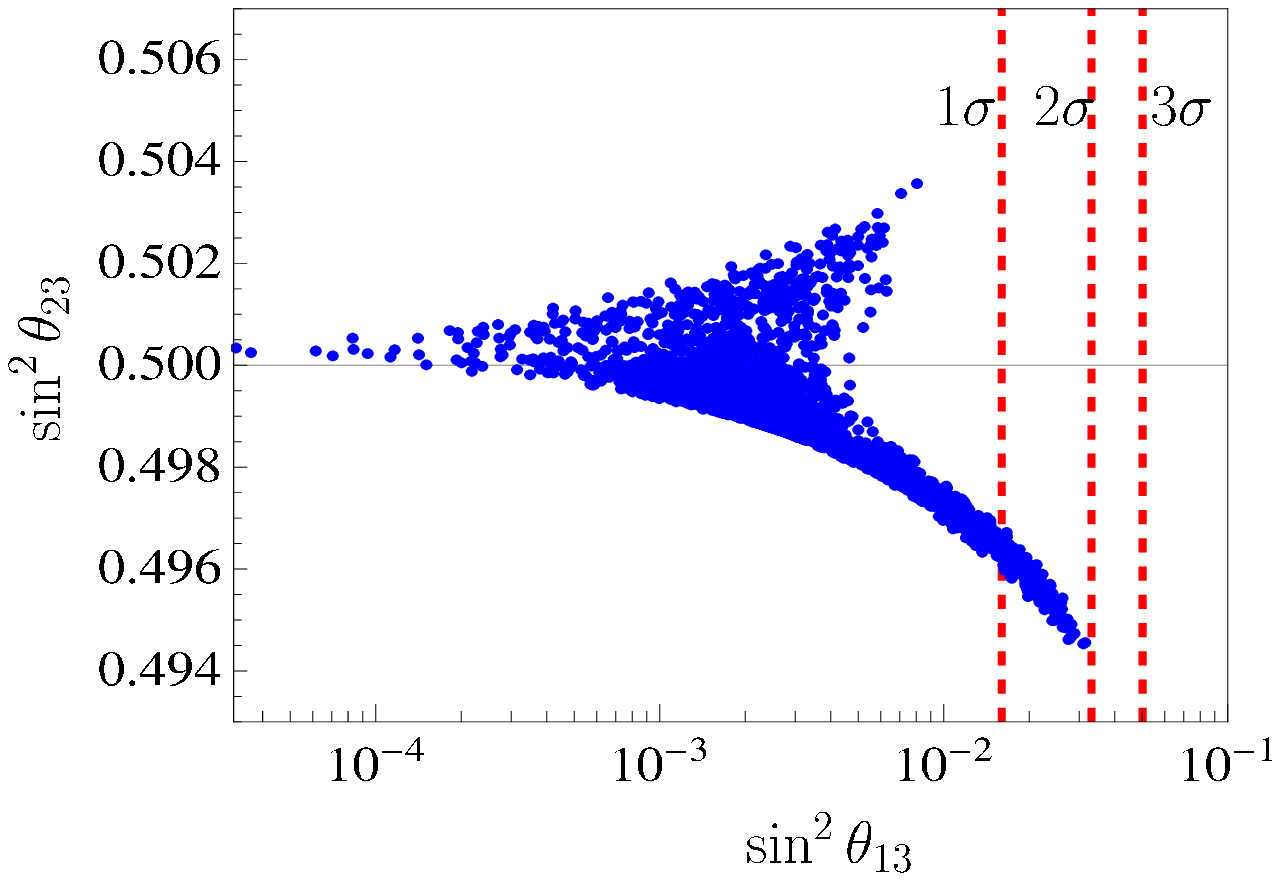}
\includegraphics[width=8cm]{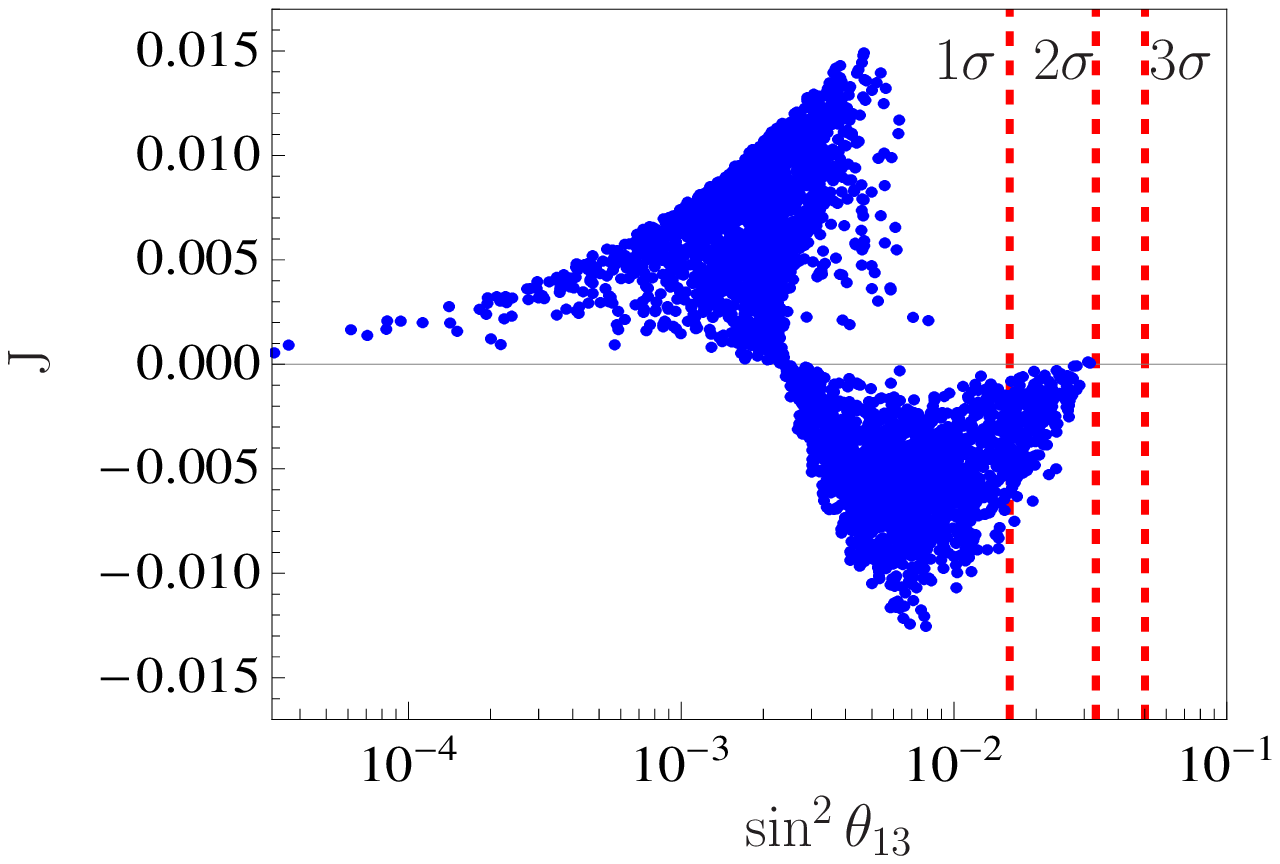}
\caption{The figure in the left side shows the allowed region for $\theta_{23}~\mbox{vs.}~\theta_{13}$. The figure in the right side shows allowed region for $J~\mbox{vs.}~\theta_{13}$.}
\label{figt}
\end{figure}

In the left side of figure \ref{figt}, we show the allowed region for the atmospheric mixing angle. The deviation from its maximal value is small. The reactor mixing angle, $\theta_{13}$, can be large.  In the right figure \ref{figt} side of we show the magnitude of the Jarlskog invariant~\cite{Jarlskog:1985cw}, $J$, of CP violation in neutrino oscillation defined as 
\be
J=Im(V_{11}V_{22} V_{12}^*V_{21}^*).
\ee
$J$ is correlated to the $\sin^2\theta_{13}$ that can be measured
in next experiments like Double Chooz~\cite{Ardellier:2006mn}.

%The Model with $\epsilon=0$, has five free paramenters, $x,~y,~\theta~\phi_{xy}$ and $\alpha$ in the neutrino sector for nine physical parameters, $\Delta m_{12}^2$, $\Delta m_{13}^2$, $m_{ee}$, three mixing angles and two Majorana phases and the Jarlskog invariant $J$, for Dirac CP violation.

%\begin{figure}[h!]
%\begin{minipage}[b]{0.5\columnwidth}
%\centering
%\includegraphics[width=8cm]{T13j}
%\caption{Jarlskog Invariant vs. reactor angle $\theta_{13}$}
%\label{fig1}
%\end{minipage} \hfill{}
%\hspace{0.5cm}
%\begin{minipage}[b]{0.5\columnwidth}
%\centering
%\includegraphics[width=8cm]{T13a}
%\caption{Correlation between the atmospheric mixing angle $\theta_{23}$ and the reactor angle $\theta_{13}$}
%\label{fig2}
%\end{minipage}
%\end{figure}

\subsection{Neutrinoless double beta decay}

Neutrinos are guaranteed to have a non-zero Majorana mass if the neutrinoless double beta decay $0\nu\beta\beta$ %violating the lepton number by two units with no missing energy, 
is observed~\cite{valle}. The $0\nu\beta\beta$ decay rate $m_{ee}$ is proportional to the $\nu_e-\nu_e$ entry of the Majorana neutrino mass matrix $M_\nu$. 
For an introduction to the phenomenology of $0\nu\beta\beta$
see for instance~\cite{Hirsch:2006tt}.

\begin{figure}[h!]
\includegraphics[width=8cm]{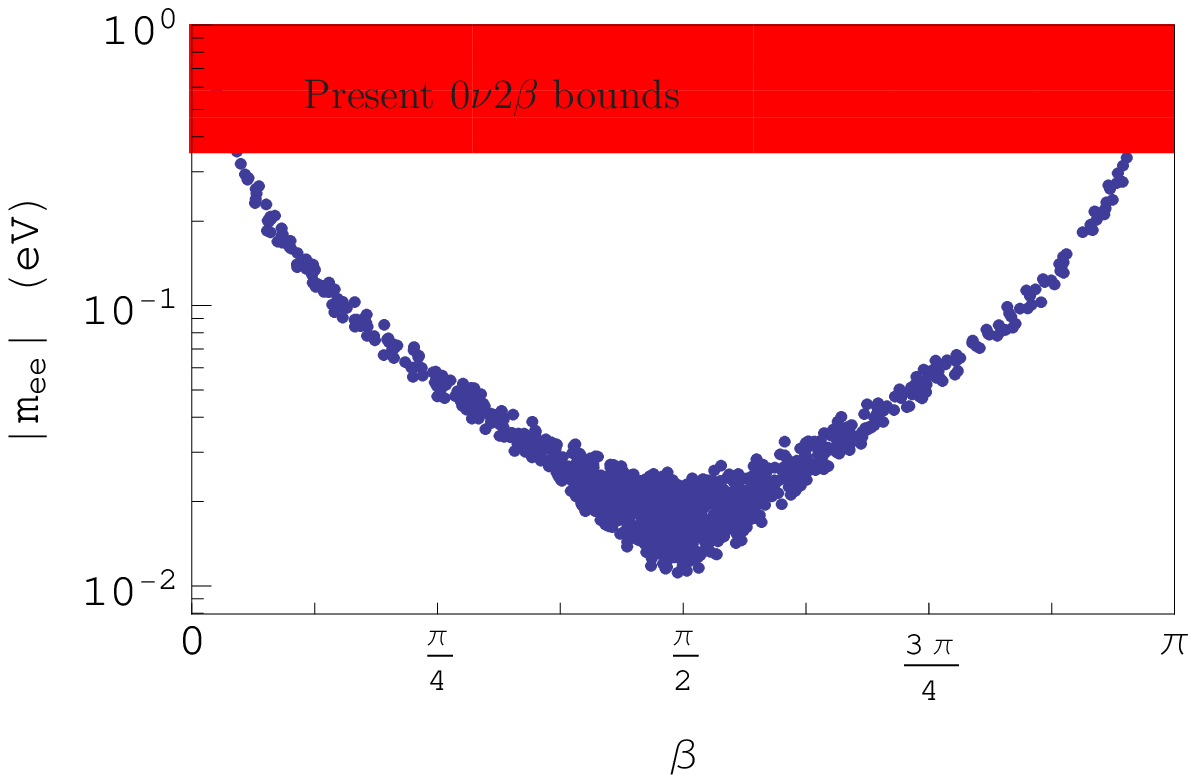}
\hspace{0.5cm}
\includegraphics[width=8cm]{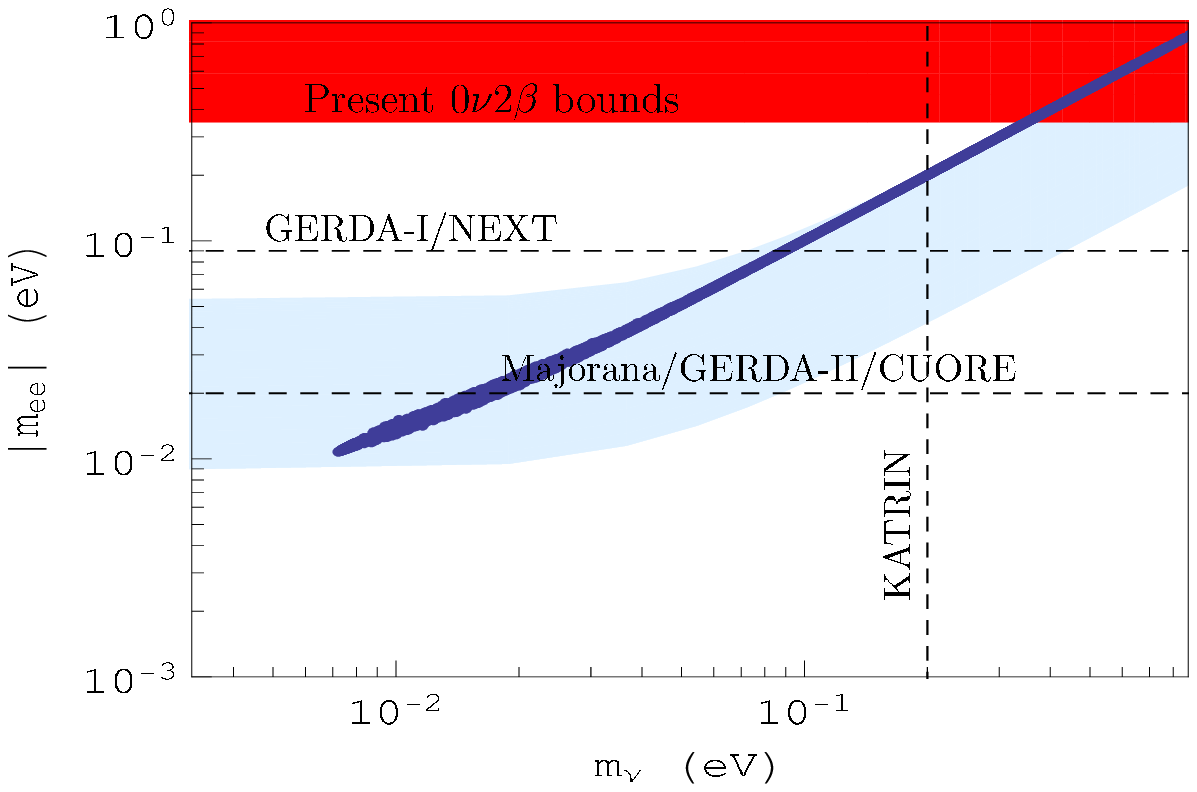}
\caption{The figure in the left side shows the allowed range for the $0\nu\beta\beta$ as function of the physical Majorana phase $\beta$. The figure in the right side shows $|m_{ee}|$ as function of the lightest neutrino mass $m_{\nu 3}$. We also present here the future experimental sensitivity~\cite{mbbd} .}
\label{fig2}
\end{figure}

We observed that the parameter $\phi_{xy}$ is related with the Majorana phase $\beta$ and the $m_{ee}$ is predicted in our model. The allowed region for the neutrino double beta decay is shown in figure~\ref{fig2}.
%In figure~\ref{fig2} 
We show the dependence of the $m_{ee}$ as function of the Majorana physical phase $\beta$. 
As can be seeing from the figure, the values for $\beta=0$ and $\beta=\pi$ of $m_{ee}$ are forbidden by the upper limits obtained by the HM collaboration.
Other models in literature excluding $\beta=0$ but not $\beta=\pi$ has been studied~\cite{Hirsch:2008rp}.
We note that $m_{\mbox{light}}> 0.008$.

\section{Conclusion}
\label{conc}

We have studied a model for lepton mixing based on a $A_4$ flavor symmetry. This constraints the model, reducing the number of free parameters with respect to the case of the Standard Model. In the scalar sector we introduce three Higgs doublets that belong to a triplet representation of $A_4$. If the vevs of the Higgs fields are assumed to be real, there are only two possible solutions: $i)$ the three vevs are all equal, or, 
$ii)$ two vevs are equal to zero. In this paper we consider the most general case with complex vevs. This solution is different from that of real vevs, it is found that one of the vevs is real and the other two are the complex conjugate one of each other, that is, $\langle H\rangle=(v_1,v,v^*)$, where $v_1$ is different from $v$ as noted also in \cite{Lavoura:2007dw}. 
This fact opens an interesting scenarios in the model building due to the extra CP phase in the Higgs sector. We studied the phenomenological implications of the neutrino masses and mixings. The charged lepton mass matrix arises only from renormalizable Yukawa interactions, while the Majorana neutrino mass matrix arises from a dimension five operator. We do not enter into details how this dimension five operator is generated.
 In order to fit the data we assumed a moderate fine tuning between the free parameters in the neutrino mass matrix. We found that the model is compatible with inverse hierarchy only. The atmospheric angle is very close to the maximal value, $\sin^2\theta_{23}\sim 0.5$ and the maximum allowed value for the reactor angle is close to the current $2\sigma$ upper bound, that is $\sin^2\theta_{13}\sim 0.03$. The solar mixing angle can be fitted in the allowed experimental range at $3\sigma$. The maximal value for the CP Jarlskog invariant is $|J|\approx 0.015$. We also found that the current $0\nu \beta\beta$ upper bound restricts the physical Majorana phase $\beta$, to be slightly different from zero and $\pi$. The power predictivity of this model in the leptonic sector make it interesting enough to make an attempt to extend this model to the quark sector. An important point is that the alignment we obtain here is incompatible with the hierarchy of the quark masses and the charged lepton masses simultaneously. A supersymetric version of this model is being invastigated~\cite{morisi2} in order to solve this problem.

\section{Acknowledgments}
This talk was based on work in collaboration with S. Morisi, and is supported 
by the Spanish MICINN under grants FPA2008-00319/FPA and MULTIDARK Consolider 
CAD2009-00064, by Prometeo/2009/091.

\end{document}